# A Vast Thin Plane of Co-rotating Dwarf Galaxies Orbiting the Andromeda Galaxy

Rodrigo A. Ibata[1], Geraint F. Lewis[2], Anthony R. Conn[3], Michael J. Irwin[4], Alan W. McConnachie[5], Scott C. Chapman[6], Michelle L. Collins[7], Mark Fardal[8], Annette M. N. Ferguson[9], Neil G. Ibata[10], A. Dougal Mackey[11], Nicolas F. Martin[1,7], Julio Navarro[12], R. Michael Rich[13], David Valls-Gabaud[14], and Lawrence M. Widrow[15]

[1]*Observatoire astronomique de Strasbourg, 11, rue de l'Université, F-67000 Strasbourg, France.*
[2]*Sydney Institute for Astronomy, School of Physics, A28, The University of Sydney, NSW 2006, Australia.*
[3]*Department of Physics and Astronomy, Macquarie University, NSW 2109, Australia.*
[4]*Institute of Astronomy, University of Cambridge, Madingley Road, Cambridge CB3 0HA, UK.*
[5]*NRC Herzberg Institute of Astrophysics, 5071 West Saanich Road, Victoria, British Columbia, V9E 2E7 Canada.*
[6]*Department of Physics and Atmospheric Science, Dalhousie University, 6310 Coburg Rd., Halifax, NS B3H 4R2 Canada*
[7]*Max-Planck-Institut für Astronomie, Königstuhl 17, 69117 Heidelberg, Germany.*
[8]*University of Massachusetts, Department of Astronomy, LGRT 619-E, 710 N. Pleasant Street, Amherst, Massachusetts 01003-9305, USA.*
[9]*Institute for Astronomy, University of Edinburgh, Royal Observatory, Blackford Hill, Edinburgh EH9 3HJ, UK.*
[10]*Lycée international des Pontonniers, 1 rue des Pontonniers, F-67000 Strasbourg, France.*
[11]*The Australian National University, Mount Stromlo Observatory, Cotter Road, Weston Creek, ACT 2611, Australia*
[12]*Department of Physics and Astronomy, University of Victoria, 3800 Finnerty Road, Victoria, British Columbia, Canada V8P 5C2.*
[13]*Department of Physics and Astronomy, University of California, Los Angeles, PAB, 430 Portola Plaza, Los Angeles, California 90095-1547, USA.*

[14]*LERMA, UMR CNRS 8112, Observatoire de Paris, 61 Avenue de l'Observatoire, 75014 Paris, France*

[15]*Department of Physics, Engineering Physics, and Astronomy Queen's University, Kingston, Ontario, Canada K7L 3N6.*

**Dwarf satellite galaxies are thought to be the remnants of the population of primordial structures that coalesced to form giant galaxies like the Milky Way[1]. An early analysis[2] noted that dwarf galaxies may not be isotropically distributed around our Galaxy, as several are correlated with streams of H I emission, and possibly form co-planar groups[3]. These suspicions are supported by recent analyses[4–7], and it has been claimed[7] that the apparently planar distribution of satellites is not predicted within standard cosmology[8], and cannot simply represent a memory of past coherent accretion. However, other studies dispute this conclusion[9–11]. Here we report the existence (99.998% significance) of a planar subgroup of satellites in the Andromeda galaxy, comprising $\approx 50\%$ of the population. The structure is vast: at least $\sim 400$ kpc in diameter, but also extremely thin, with a perpendicular scatter $< 14.1$ kpc (99% confidence). Radial velocity measurements[12–15] reveal that the satellites in this structure have the same sense of rotation about their host. This finding shows conclusively that substantial numbers of dwarf satellite galaxies share the same dynamical orbital properties and direction of angular momentum, a new insight for our understanding of the origin of these most dark matter dominated of galaxies. Intriguingly, the plane we identify is approximately aligned with the pole of the Milky Way's disk and is co-planar with the Milky Way to Andromeda position vector. The existence of such extensive coherent kinematic structures within the halos of massive galaxies is a fact that must be explained within the framework of galaxy formation and cosmology.**

We undertook the Pan-Andromeda Archaeological Survey[16] (PAndAS) to obtain a large-scale panorama of the halo of the Andromeda galaxy (M31), a view that is not available to us for the Milky Way. This Canada-France-Hawaii Telescope survey imaged $\sim 400$ square degrees around M31, the only other giant galaxy in the Local Group apart from the Milky Way. Stellar objects are detected out to a projected distance of $\sim 150$ kpc from M31, and $\sim 50$ kpc from M33, the most massive satellite of M31. The data reveal a substantial population of dwarf spheroidal galaxies that accompany Andromeda[17].

The distances to the dwarf galaxies can be estimated by measuring the magnitude of the Tip of the Red Giant Branch (TRGB)[18]. Improving on earlier methods, we have developed a Bayesian approach that yields the probability distribution function (PDF) for the distance to each individual satellite[19]. In this way we now have access, for the first time, to homogenous distance measurements (typical uncertainties 20–50 kpc) to the 27 dwarf galaxies (filled circles in Figure 1) visible within the PAndAS survey area[20], that lie beyond the central 2°.5.

In Figure 2 these distance measurements are used to calculate the sky positions of the homogenous sample of 27 dwarf galaxies as they would appear from the centre of the Andromeda galaxy. Visually, there appears to be a correlation close to a particular great circle (red line): this suggests that there is a plane, centred on M31, around which a subsample of the satellites have small scatter. This is confirmed by the Monte Carlo analysis presented in the Supplementary Information, where we show that the probability of the alignment of the sub-sample of $n_{sub} = 15$ satellites marked red in Figures 1 and 2 occurring at random is 0.13% (see Figure S1).

Following this discovery, we sought to investigate whether the sub-sample displayed any kinematic coherence. The radial velocity of each satellite is shown in Figure 3, corrected for the bulk motion of the Andromeda system towards us: what is immediately striking is that 13 of the 15 satellites possess coherent rotational motion, such that the southern satellites are approaching us with respect to M31, while the northern satellites recede away from us with respect to their host galaxy.

The probability that 13 or more out of 15 objects should share the same sense of rotation is 1.4% (allowing for right-handed or left-handed rotation). Thus the kinematic information confirms the spatial correlation initially suspected from a visual inspection of Figure 2. The total significance of the planar structure is approximately 99.998%.

Thus we conclude that we have detected, with very high confidence, a coherent planar structure of 13 satellites with a root-mean-square (rms) thickness of $12.6 \pm 0.6$ kpc ($< 14.1$ kpc at 99% confidence), that co-rotate around M31 with a (right-handed) axis of rotation that points approximately East. The three-dimensional configuration can be assessed visually in Figure 3. The extent of the structure is gigantic, over 400 kpc along the line of sight and nearly 300 kpc North-South. Indeed, since And XIV and Cas II lie at the Southern and Northern limits of the PAndAS survey, respectively, it is quite probable that additional (faint) satellites belonging to this structure are waiting to be found just outside the PAndAS footprint. Although huge in extent, the structure appears to be lopsided, with most satellites populating the side of the halo of M31 nearest the Milky Way. The completeness analysis we have undertaken shows that this configuration is not due to a lowered detection sensitivity at large distance, but reflects a true paucity of satellites in the more distant halo hemisphere[21].

The existence of this structure had been hinted at in earlier work[22], thanks to the planar alignment we report on here being viewed nearly edge-on, although the grouping could not be shown to be statistically significant from the information available at that time. Other previously-claimed alignments do not match the present plane, although they share some of the member galaxies: the most significant alignment of Koch & Grebel[23] has a pole 45° away from that found here, and the (tentative) poles of the Metz et al.[4] configuration are 23°.4 away, while those of a later contribution[6] have poles 34°.1 and 25°.4 distant from ours. Without the increased sample size, reliable three-dimensional positions and radial velocities, and most importantly, a spatially unbiassed selection function resulting from the homogenous panoramic coverage of PAndAS, the nature, properties and conclusive statistical significance of the present structure could not be inferred.

The present detection proves that in some giant galaxies, a significant fraction of the population of dwarf satellite galaxies, in this case $\approx$ 50% (13 out of 27 over the homogeneously-surveyed PAndAS area), are aligned in coherent planar structures, sharing the same direction of angular momentum. The Milky Way is the only other giant galaxy where we have access to high-quality three-dimensional positional data, and the existence of a similar structure around our Galaxy is strongly suggested by current data[2, 5, 7]. The implications for the origin and dynamical history of dwarf galaxies are profound. It also has a strong bearing on the analyses of dark matter in these darkest of galaxies, since one cannot now justifiably assume such objects to have evolved in dynamical isolation.

Intriguingly, the Milky Way lies within 1° of the plane reported here, the pole of the plane and the pole of the Milky Way's disk are approximately perpendicular (81°), and furthermore this plane is approximately perpendicular to the claimed plane of satellites that surrounds the Galaxy (since its pole points approximately towards Andromeda[7] within the uncertainties). While these alignments may be chance occurrences, it is nevertheless essential information about the structure of the nearby Universe that will have to be taken into account in future simulations aimed at modelling the dynamical formation history of the Local Group.

The formation of this structure around M31 poses a difficult puzzle. For discussion, we envisage two broad classes of possible explanations: accretion or in-situ formation. In either type of model, the small scatter of the satellites out of the plane is a challenge to explain, even though the orbital timescales for the satellites are long (∼ 5 Gyr for satellites at 150 kpc). All the galaxies in the plane are known to have old, evolved, stellar populations, and so in-situ formation would additionally imply that the structure is ancient.

In an accretion scenario, the dynamical coherence points to an origin in a single accretion of a group of dwarf galaxies. However, the spatial extent of the progenitor group would have to be broadly equal to or smaller than the current plane thickness (< 14.1 kpc), yet no such groups are known. Interpreting the coherent rotation as a result of our viewing perspective[24] requires a bulk tangential velocity for the in-falling group of order ∼ 1000 km s$^{-1}$, which seems unphysically high. A further possibility is that we are witnessing accretion along filamentary structures that are fortuitously aligned. In-situ formation may be possible if the planar satellite galaxies formed like tidal-dwarf galaxies (TDG) in ancient gas-rich galaxy mergers[7], but then the dwarf galaxies should be essentially devoid of dark matter. If the planar M31 dwarfs are dynamically-relaxed, the absence of dark matter would be greatly at odds with inferences from detailed observations[25] of Milky Way satellites assuming the standard theory of gravity. An alternative possibility is that gas was accreted preferentially onto dark matter sub-halos that were already orbiting in this particular plane, but then the origin of the plane of sub-haloes would still require explanation. We conclude that it remains to be seen whether galaxy formation models within the current cosmological framework can explain the existence of this vast, thin, rotating structure of dwarf galaxies within the halo of our nearest giant galactic neighbour.

**Acknowledgements**   We thank the staff of the Canada-France-Hawaii Telescope for taking the PAndAS data, and for their continued support throughout the project. We would like to thank one of our referees, Prof. B. Tully, for pointing out that IC 1613 could also be associated to the planar structure. R.A.I. and D.V.G. gratefully acknowledges support from the Agence Nationale de la Recherche though the grant POMMME, and would like to thank Benoit Famaey for insightful discussions. G.F.L thanks the Australian research council for support through his Future Fellowship and Discovery Project. Based on observations obtained with MegaPrime/MegaCam, a joint project of CFHT and CEA/DAPNIA, at the Canada-France-Hawaii Telescope (CFHT) which is operated by the National Research Council (NRC) of Canada, the Institut National des Sciences de l'Univers of the Centre National de la Recherche Scientifique (CNRS) of France, and the University of Hawaii. Some of the data presented herein were obtained at the W.M. Keck Observatory, which is operated as a scientific partnership among the California Institute of Technology, the University of California and the National Aeronautics and Space Administration. The Observatory was made possible by the generous financial support of the W.M. Keck Foundation.

**Competing Interests**   The authors have no competing financial interests.

**Author contributions**   All authors assisted in the development and writing of the paper. In addition, the structural and kinematic properties of the dwarf population, and the significance of the Andromeda plane were determined by R.A.I., G.F.L. and A.R.C., based on distances determined by the same group (as part of the PhD research of A.R.C.). In addition, A.W.M. is the Principal Investigator of PAndAS; M.J.I. and R.A.I led the data processing effort; R.A.I. was the Principal Investigator of an earlier CFHT MegaPrime/MegaCam survey, which PAndAS builds on (which included S.C.C., A.M.N.F., M.J.I., G.F.L., N.F.M. and A.W.M.). R.M.R. is Principal Investigator of the spectroscopic followup with the Keck Telescope. M.L.C. and S.C.C. led the analysis of the kinematic determination of the dwarf population, and N.F.M. led the detection of the dwarf population from PAndAS data. N.G.I. performed the initial analysis of the satellite kinematics.

**Correspondence**   Reprints and permissions information is available at www.nature.com/reprints. Correspondence and requests for materials should be addressed to R.A.I. (rodrigo.ibata@astro.unistra.fr).

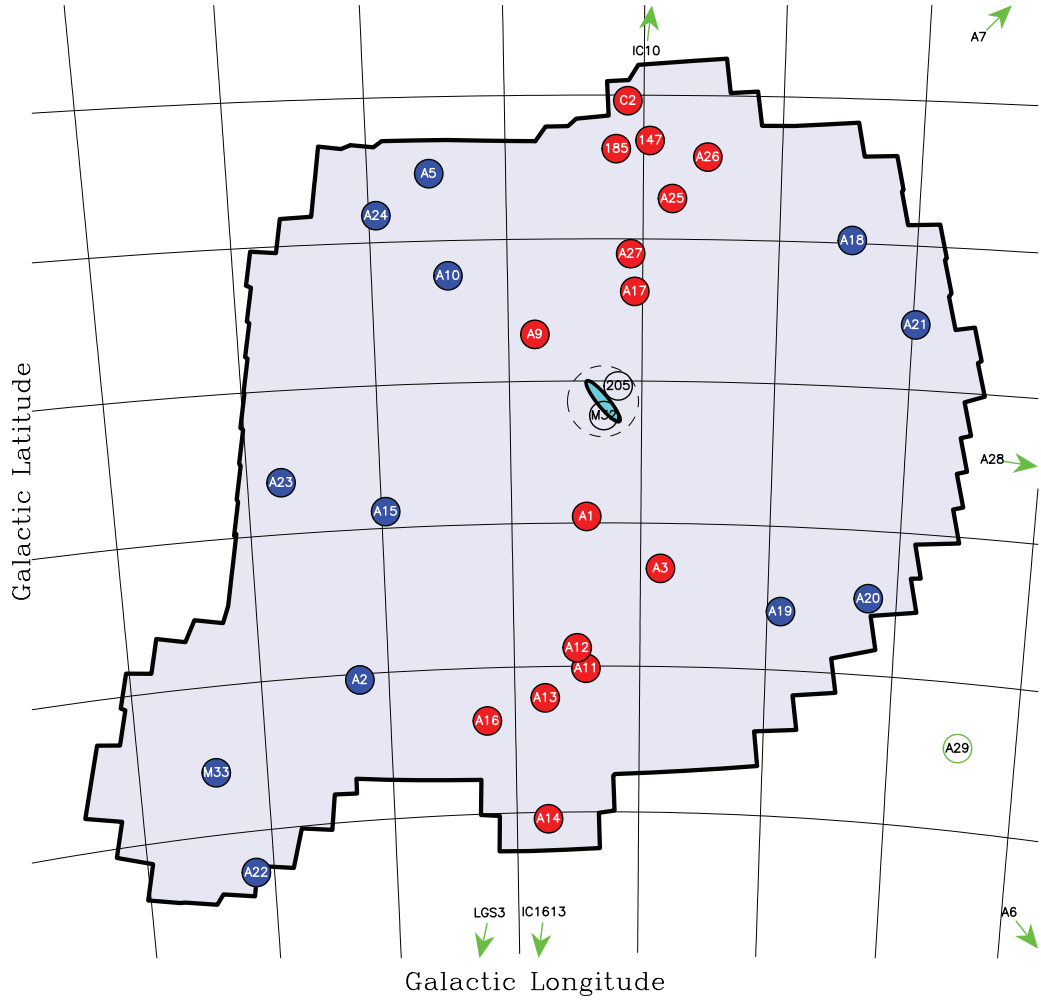

Figure 1: Map of the Andromeda satellite system. The homogenous PAndAS survey (irregular polygon) provides the source catalogue for the detections and distance measurements of the 27 satellite galaxies[20] (filled circles) used in this study. Near M31 (ellipse), the high background hampers the detection of new satellites and precludes reliable distance measurements for M32 and NGC 205 (black open circles); we therefore exclude the region inside 2°.5 (dashed circle) from the analysis. The seven satellites known outside the PandAS area (green circles/arrows) constitute a heterogenous sample, discovered in various surveys with non-uniform spatial coverage, and their distances are not measured in the same homogenous way. Since a reliable spatial analysis requires a dataset with homogenous selection criteria, we do not include these objects in the sample either. The analysis shows that satellites marked red are confined to a highly planar structure. Note that this structure is approximately perpendicular to lines of constant Galactic latitude, so it is therefore aligned approximately perpendicular to the Milky Way's disk (the grid squares are 4° × 4°).

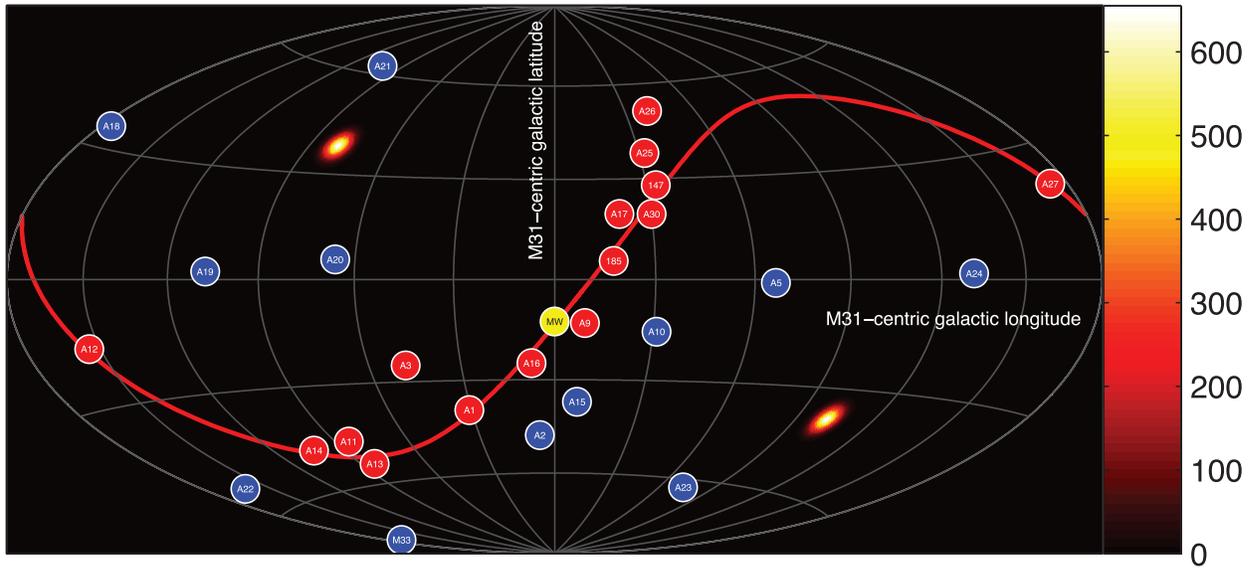

Figure 2: Satellite galaxy positions as viewed from Andromeda. The (Aitoff-Hammer) projection shows the sample of 27 satellites[20] (filled circles from Figure 1) as they would be seen from the centre of the Andromeda galaxy. In these coordinates the disk of Andromeda lies along the equator. The background image represents the probability density function of the poles derived from $10^5$ iterations of resampling the 27 satellites from their distance PDFs, and finding the plane of lowest rms from a sub-sample of 15 (the colour wedge shows the relative probability of the poles). A clear narrow peak at ($l_{M31} = 100°.9 \pm 0°.9$, $b_{M31} = -38°.2 \pm 1°.4$) highlights the small uncertainty in the best-fit plane. The solid red line, which passes within less than 1° of the position of the Milky Way (yellow circle), represents the plane corresponding to this best pole location.

Figure 3: (**Next page**) Three-dimensional view of the planar, rotating structure **(can be visualised and manipulated in 3D with Adobe Reader)**. The coordinate system is such that the $z$ direction is parallel to the vector pointing from the Milky-Way to M31, $x$ increases eastwards and $y$ northwards. Only the radial component of velocity of the satellites is measured, these are shown as vectors pointing either towards or away from the Milky Way. As in Figs. 1 and 2, red spheres mark the planar satellites, while blue spheres represent the "normal" population. The coherent kinematic behaviour of the spatially very thin structure (red) is clearly apparent viewed from the $y$–$z$ plane. With the exception of And XIII and And XXVII, the satellites in the planar structure that lie to the North of M31 recede from us, while those to the South approach us; this property strongly suggests rotation. Our velocity measurements[15] (supplemented by literature values[14]), have very small uncertainties, typically < 5 km s$^{-1}$. The irregular green polygon shows the PAndAS survey area, the white circle indicates a projected radius of 150 kpc at the distance of M31, and the white arrow marks a velocity scale of 100 km s$^{-1}$. (And XXVII is not shown in this diagram as its most likely distance is 476 kpc behind M31). This figure is interactive (allowing the reader to change the magnification and viewing angle), and was constructed with the S2PLOT programming library[26].

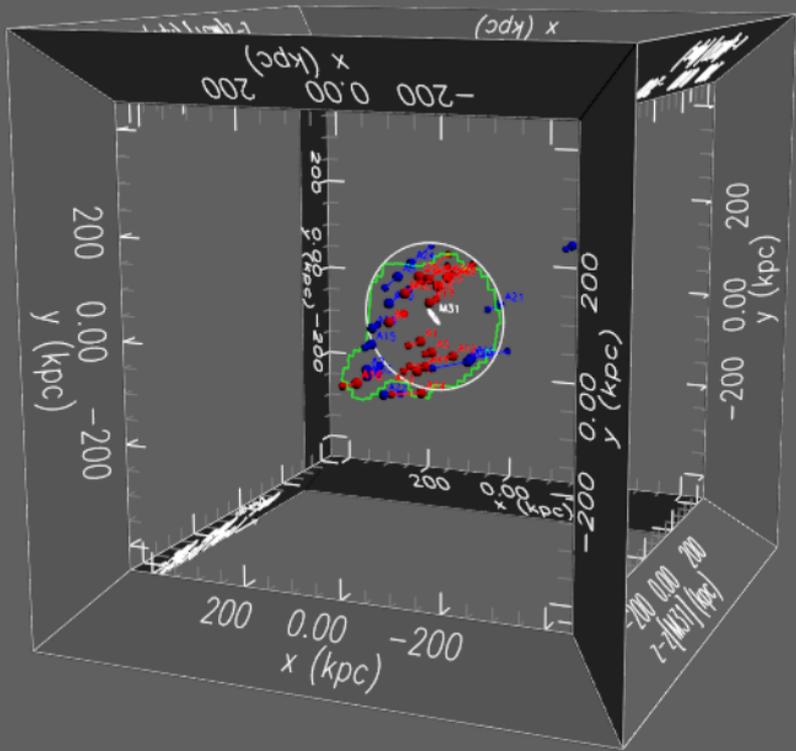

# Supplementary Information

## 1  Monte Carlo tests for statistical significance

To determine the best plane that fits the data it is necessary to take the distance uncertainties into account. This is accomplished as follows: first, we generate a set of 27 (three-dimensional) positions, by randomly drawing from the distance PDFs of each satellite. Subsequently, we find the plane that has the lowest root-mean-square (rms) distance to any sub-sample of $n_{sub} = 15$ satellites. By repeating 1000 times the procedure of drawing 27 satellites from their distance PDFs and calculating the lowest rms plane, we obtain a probability density function for the root mean square thickness of a possible planar sub-structure of $n_{sub} = 15$ satellites given the data. The distribution is approximately Gaussian with $\sigma = 0.6$ kpc, and mean 12.6 kpc. The 15 satellites that are closest to the lowest rms plane are marked with red circles in Figures 1 and 2.

The next question we need to assess is: given the M31-centric distance distribution to the satellites, what is the chance that they could be arranged at random to form a planar structure with equal or lower rms? To answer this question we performed a careful Monte Carlo simulation using random realisations of satellite configurations that are isotropically-distributed in the Andromedan sky, but preserve the same distribution of radial (not projected) distances to the centre of M31 as the observed one. To achieve this, an artificial satellite is generated by selecting a satellite at random from the real set, giving it a random 3-dimensional orientation with respect to Andromeda, and assigning the same line of sight distance PDF as the real satellite has (although shifted to the new distance). If the resulting artificial satellite is located outside of the PAndAS area, or within a projected distance of 2°.5, it is discarded. In this way we draw a random sample of 27 artificial satellites (with replacement) which we process in an identical way as we processed the real sample: creating a new set of 1000 rms values, from which we construct a probability density function for the rms for that particular sample of 27 artificial satellites. The whole procedure is repeated $10^5$ times. The corresponding histogram of mean rms values in the artificial satellite samples is shown in Figure S1. We find that a mean rms scatter of 12.6 kpc or less occurs with probability 0.13%.

Thus the sample of 15 satellites that is closely aligned on the great circle in Figure 2 (red circles) is indeed a highly surprising and statistically significant detection.

The above choice of $n_{sub} = 15$ members for the sub-sample size was made by examining the effect that this parameter has on the rms thickness and on the position of the best-fit plane. We find that for $5 \leq n_{sub} \leq 15$ the position of the pole to the lowest rms plane remains very stable, while the plane rms thickness increases only slowly with increasing $n_{sub}$. However, for values of $n_{sub} \geq 16$ the plane rms increases rapidly, and also there is no well-defined plane solution, unlike the tight solution for $n_{sub} = 15$ displayed as the background image in Figure 2. A slightly higher significance is derived by choosing an $n_{sat}$ value of 13 or 14, as the smaller rms scatter about the plane more than compensates for the lower statistics in the sub-sample.

## 2  The planar satellites

The co-rotating satellites are: And I, And III, And IX, And XI, And XII, And XIV, And XVI, And XVII, And XXV, And XXVI, Cas II, NGC 147 and NGC 185. The two satellites that do not partake in this motion are And XIII and And XXVII; it seems likely that these are interlopers from the normal non-planar population, although they may nevertheless be members of the planar subgroup if the population has significant velocity dispersion. Figure 1 shows that of the known M31 satellite galaxies, M32, NGC 205, LGS 3, IC 10 and IC 1613 (the latter three are situated up to 40 degrees from M31 in the directions indicated) also lie along the same axis as the red-coloured objects from our sample, and published distance estimates[28] are consistent with all five satellites being at the same distance as their host within the uncertainties. However, only LGS 3, IC 1613 and NGC 205 share the same sense of rotation as the 13 planar satellites listed above, and may plausibly be associated with the structure.

No obvious differences in the physical properties (metallicity, stellar populations or velocity dispersion) were found between the 13 kinematically-coherent coplanar satellites and the remainder of the sample.

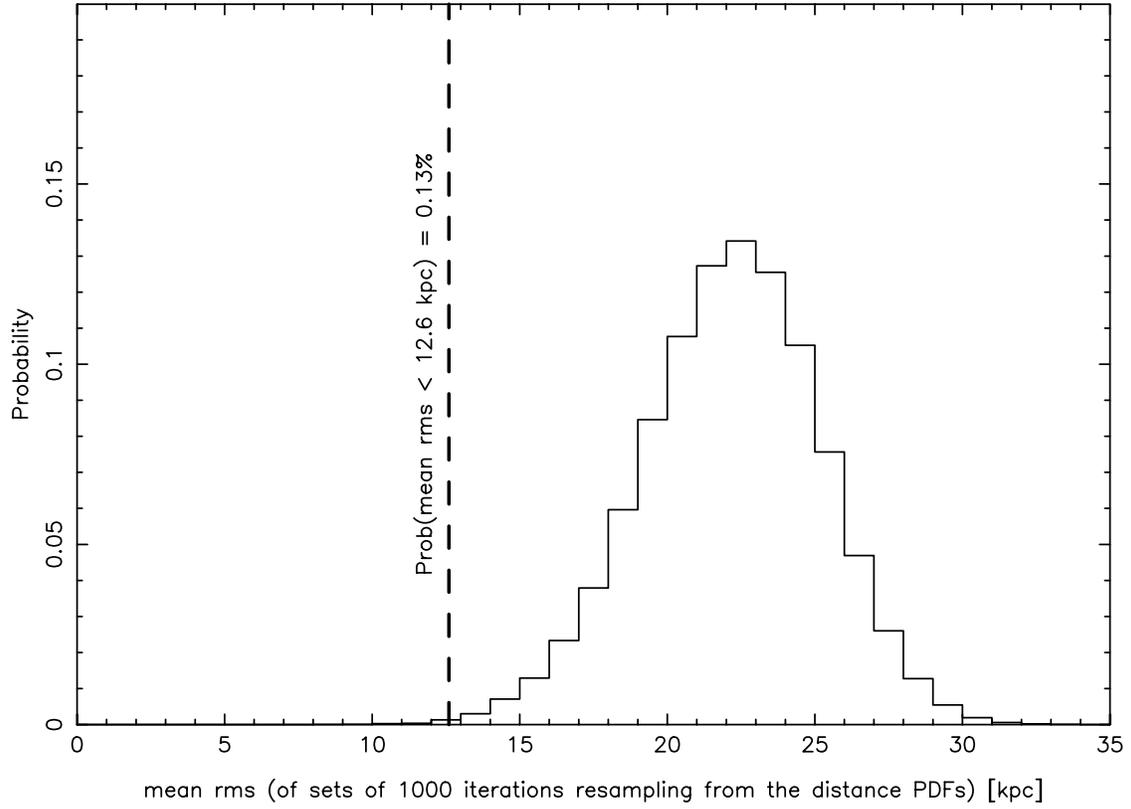

Figure S1: Statistical significance of the spatial alignment. The histogram presents the distribution in $10^5$ random trials of the average root-mean-square distances from a best-fit plane of sub-samples of 15 satellites. The average rms value of 12.6 kpc, derived from the real configuration, is extremely rare, occurring with probability 0.13% in random realisations. This demonstrates that the observed spatial alignment of dwarf galaxies is very unlikely to be a random coincidence. If we had chosen a smaller sub-sample size, the chance alignment becomes even more extreme: e.g. 0.06% for $n_{sub} = 13$.